\documentclass[a4paper,11pt]{article}

\usepackage{fullpage}
\usepackage{physics}
\usepackage{amssymb}
\usepackage{graphicx}
\usepackage{hyperref}
\DeclareGraphicsExtensions{.pdf}

\newcommand{\microns}{\ensuremath{\mathrm{\mu m}}}%

\begin{document}

\title{Tracking charged particles in the zero curvature limit}
\author{J. Alcaraz Maestre \\ CIEMAT-Madrid}
\date{CIEMAT Technical Report 1487
\footnote{ISBN/ISSN: 2695-8864, NIPO: 832-21-005-X, CIEMAT, September 2021, \url{https://cpage.mpr.gob.es/}}
} 

\maketitle

\begin{abstract}
In this report we discuss appropriate strategies for the tracking of charged particles in
the limit of zero curvature. The suggested approach avoids special treatments and precision
issues that frequently arise in that limit. We provide explicit expressions for transport,
refitting and vertexing in regions where magnetic field inhomogeneities or detector
interaction effects can be approximately ignored. 
\end{abstract}

\section{Introduction \label{sec:introduction}}

The linear relation between the transverse momentum $p_T$ of a charged particle and the \
inverse of its curvature $C$ in the presence of a magnetic field constitutes the basis for
the determination of momenta in particle physics. This relation is typically exploited by
tracking and analyzing the ionization deposits along the trajectory of the particle.
Sophisticated tracking techniques in past and present experiments have resulted into major
physical results and discoveries. Getting sufficiently precise measurements will become
particularly challenging at future colliders with multi-TeV center-of-mass energies, due to
the much lower expected curvatures. In this context, ultra-precise detectors with $\lesssim
10~\microns$ of position and alignment accuracies over distances of meters, able to support
high occupancies of hundreds or even thousands of tracks per event will be required. 

The zero curvature limit ($C\rightarrow 0$) or, equivalently, the transverse momentum limit
($p_T\rightarrow \infty$) is always a potential source of problems for tracking algorithms,
due to presence of divergent operations in many of the used equations. A
dangerous approach is to consider $C=0$ as a special, exceptional case. The problem occurs
in practice when the curvature gets smaller or comparable to the computer precision used in
the calculations, $\mathcal{C}_{prec}$. Apart from the inherent complications in the code, 
the scheme is intrinsically limited because it abandons by construction the possibility of 
providing sensible measurements for tracks with $C<\mathcal{C}_{prec}$. 

This report discusses tracking in the zero curvature limit, based on some past notes
written in the running period of the L3 detector at LEP~\cite{l3note}. It adopts the basic
strategy of avoiding the use of any curvature threshold in a systematic way. For that, we
work at all times with relatively simple expressions that do not present divergences in the
$C\rightarrow 0$ limit. This turns out to be optimal to obtain precise measurements at high
momenta. The resulting equations were successfully employed already in past tracking and
detector alignment algorithms, starting with the L3 detector at
LEP~\cite{L3,l3note,smd,tec}. 

Only tracking in homogeneous magnetic fields is considered, but one has to keep in mind
that all inhomogeneous cases are decomposed in practice into a series of small propagation
steps where the present formulae are always applicable. Material effects, which are more
frequently discussed in connection with filtering techniques~\cite{fruhwirth}, do not have
an influence on the choice of the parametrization near the collision point. Specific
approaches to the case of slightly inhomogeneous magnetic fields are discussed for instance
in References~\cite{Saxon,amspaper}. 

The document is organized as follows. Section~\ref{sec:conventions} introduces the
``perigee'' parametrization that will be used throughout most of the report. It also
connects it with other equivalent conventions used in the literature. The propagation of 
track parameters and related uncertainties is discussed in 
sections\ref{sec:answers}~\ref{sec:changeref}, including specific subsections for the cases 
of propagation to cylinder and plane surfaces (sections~\ref{subsec:cylinder} 
and~\ref{subsec:plane}). Sections~\ref{sec:includepoint} and~\ref{sec:fixpoint} address 
the typical problem of track refitting when new measurements are
present, which is of utmost importance in alignment, fitting and filtering techniques in
general. Finally, sections~\ref{sec:vertex_xy} and~\ref{sec:vertex_sz} discuss vertex
determination in a simple but precise way. 

\section{Conventions \label{sec:conventions}}

\begin{figure}[ht]
\begin{center}
    \includegraphics[width=0.9\linewidth]{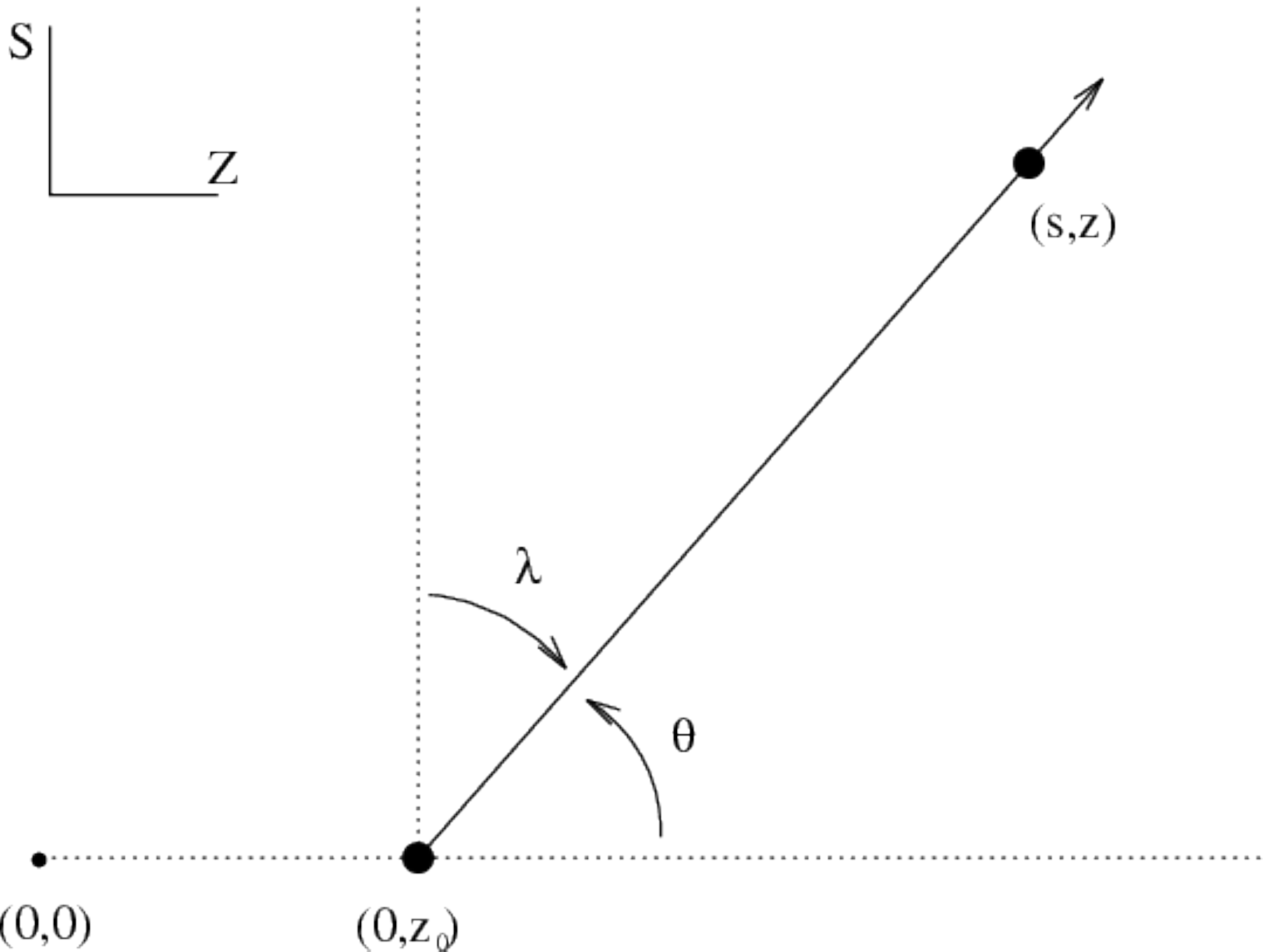}
\end{center}

\caption{Following the adopted convention, the projection of the track
on the SZ plane will be a straight line: $z=z_0 + s \tan\lambda$.  The
relevant parameters are shown in the Figure. Note that the variable $s$
at a point $(x,y,z)$ is equal to the arc length in the XY plane between
$(x_0,y_0)$ to $(x,y)$. This also implies that $s=0$ at $z=z_0$.}

\label{szfig}
\end{figure}

A helix is the most general trajectory followed by a charged particle in a homogeneous,
constant in time, magnetic field region in vacuum. We assume that the direction of the
magnetic field coincides with the positive direction of the Z axis in the reference system.
The trajectory can thus be visualized in XY as a circumference (Figure~\ref{circlefig}).
The Z displacement between two positions in the trajectory is proportional to the arc
length that is described in the XY plane, leading to the concept of a ``straight line in
the SZ plane'', where S identifies the variable associated to the arc length
(Figure~\ref{szfig}). Unless otherwise stated, equations assume that distances are given in
meters, magnetic field strengths in Teslas, charge in units of the positron charge and
momenta in GeV/$c$. We will consider a standard decomposition of the local momentum vector,
$\hat{p}$, in spherical coordinates: $\hat{p} \equiv
(p\sin\theta\cos\phi,p\sin\theta\sin\phi,p\cos\theta)$, where $p$ is its magnitude and
$\phi$ and $\theta$ are the associated azimuthal and polar angles. Since the magnetic field
is homogeneous and does not change with time, $p$ and $\theta$ are constants of motion. The
angles are chosen such that: $\phi \in (-\pi,\pi]$ and $\theta \in [0,\pi]$. 

  In the XY plane the trajectory can be fully described by a reference point, 
$(x_r,y_r)$, and three parameters, $C$, $\phi_0$ and $\delta$ 
(Figure~\ref{circlefig}):

\begin{figure}[ht]
\begin{center}
    \includegraphics[width=0.9\linewidth]{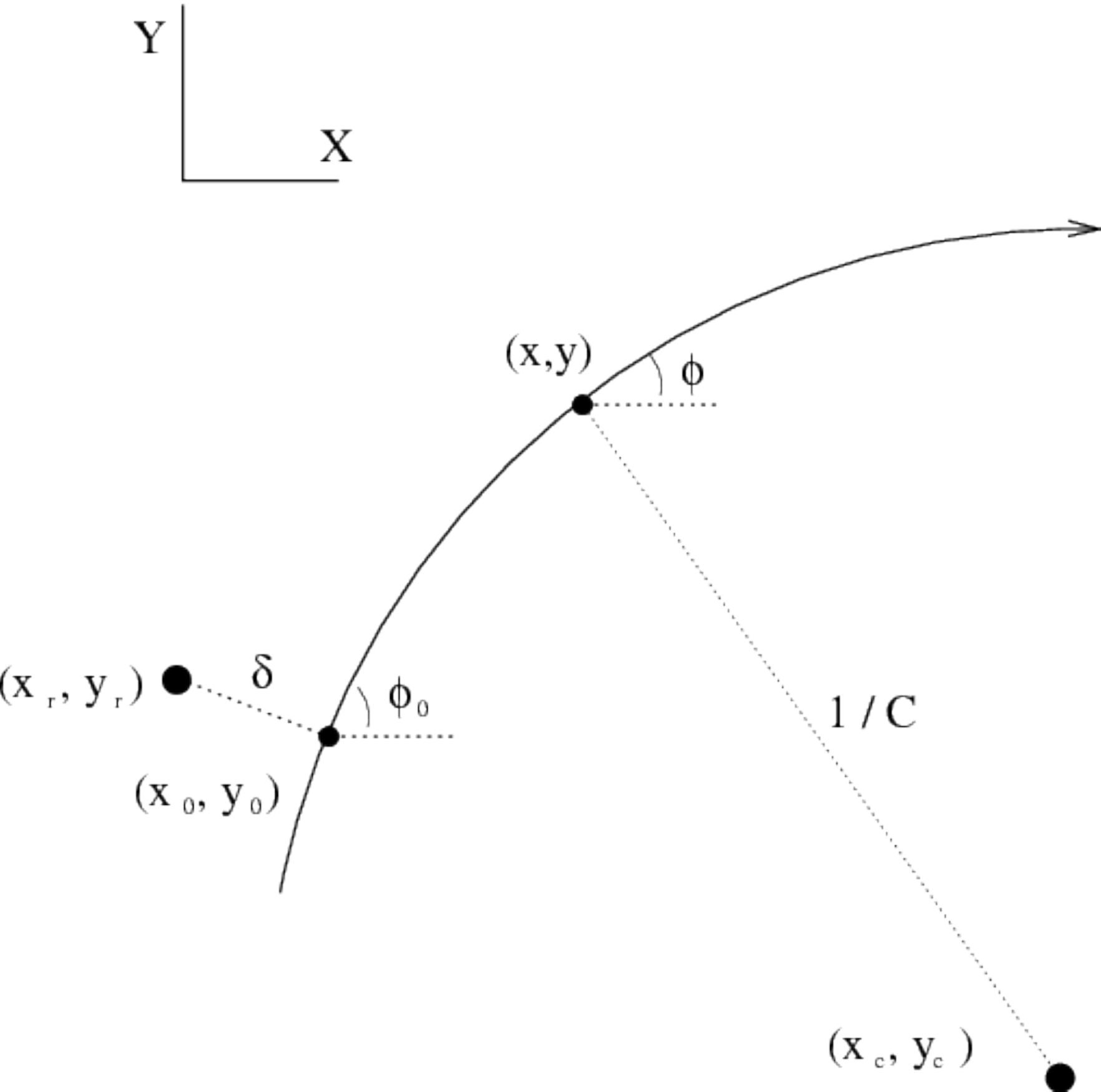}
\end{center}

\caption{ The projection of the track helix on the XY plane will be a
circumference. The parameters shown in the figure are described in the
text.}

\vfill

\label{circlefig}
\end{figure}

\begin{itemize}

\item $C$: the curvature of the track. It is related with 
      the absolute value of the transverse momentum $p_T$ via the 
      Equation: $C=0.29979~\mathrm{B q}/p_T$, where B is the magnetic 
      field strength and $q$ the particle charge. The curvature is 
      positive (negative) if the particle has a positive (negative) 
      charge. A positive (negative) curvature identifies a clockwise 
      (anti-clockwise) rotation of the particle in the XY coordinate 
      system as a function of time.
\item $\phi_0$: the azimuthal angle of the momentum vector
      at the position of closest approach to the reference point. 
\item $\delta$: the distance of closest approach to the reference 
      point. This is a signed parameter, with a convention such that 
      the coordinates of closest approach $(x_0,y_0)$ are given by: 
      \begin{eqnarray} 
        \left.
        \begin{array}{c}
         x_0 = x_r - \delta \sin\phi_0 \\
         y_0 = y_r + \delta \cos\phi_0 \\
        \end{array} \right\} & \Longrightarrow &
         ~\delta = -(x_0-x_r) \sin\phi_0 + (y_0-y_r) \cos\phi_0
      \end{eqnarray}

\end{itemize}

The trajectory in the SZ plane is simply described by a straight line:
\begin{eqnarray}
z & = & z_0 + s~\tan\lambda
\end{eqnarray}

\noindent where $s$ is the arc length traversed when the particle moves
from $(x_0,y_0)$ to $(x,y)$ and the parameters $\tan\lambda$ and $z_0$
have a simple interpretation (Figure~\ref{szfig}):

\begin{itemize}

\item $\tan\lambda$: the slope in the SZ plane, $dz/ds$. It 
      is directly related with the polar angle $\theta$ by:
      $\tan\lambda = 1/\tan\theta$.

\item $z_0$: the Z position when the particle is at the distance 
      of closest approach in the XY plane.

\end{itemize}

Let us note that the arc length $s$ can be either positive or negative,
depending on whether $(x,y)$ corresponds to a position occurring before
or after $(x_0,y_0)$ in time.

In summary, in this ``perigee'' convention, a track is fully defined by
a reference point $(x_r,y_r)$ and the set of parameters
$(C,\phi_0,\delta,z_0,\tan\lambda)$. This is the convention used for
instance by the CMS~\cite{CMS} and L3~\cite{L3} experiments to describe
track parameters and uncertainties close to the interaction
point~\footnote{The L3 parametrization was only differing by a change
in the sign convention for the $\delta$ parameter.}. According to the
previous discussion, $C$ and $\tan\lambda$ are constant over the
trajectory. 

\subsection{Alternative parametrizations \label{subsec:other_params}}

  There are equivalent parametrizations in the literature, 
  already used in past or existing high-energy physics experiments. They 
differ slightly from the one employed in this report:

\begin{itemize}

      \item An alternative perigee parametrization~\cite{Sijin}, rather
similar to our choice. The most relevant difference is the use of the parameter $\theta$
instead of $\tan\lambda$. Both parameters are constant in the case of homogeneous fields,
but $\tan\lambda$ is the slope of the straight line fit in the SZ plane, a technically 
natural coefficient for detectors embedded in a solenoidal field. Explicitly, 
this alternative choice employs the parameters $\rho$, $\phi_p$, $\epsilon$, $\theta$ 
and $z_p$, which are connected with our parameters through the relations: $\rho=-C$, $\phi_p=\phi_0$, $\epsilon=-\delta$, $\cot\theta=\tan\lambda$, $z_p=z_0$. 

      \item A variation of the previous ``perigee''
parametrization~\cite{Sijin} is the one described for instance in
Reference~\cite{Salzburger}, which is used in ATLAS~\cite{ATLAS}. It basically substitutes
the curvature $C$ by the inverse of the transverse rigidity, $\mathrm{q}/p_T$. The relation
between the curvature and $\mathrm{q}/p_T$ is: $\mathrm{q}/p_T =
(C~\cos\lambda)/(0.29979~\mathrm{B})$. Note that this parametrization requires the a priori
knowledge of the magnetic field strength, something that is not intrinsically necessary to
reconstruct tracks in regions with homogeneous magnetic fields and negligible interactions
with the medium. 

      \item The curvilinear parametrization~\cite{GEANE} adopts a local
description of the track with the parameters $(q/p, \phi, x_\perp, \lambda, z_\perp)$. Here
$x_\perp$ and $z_\perp$ quantify displacements in a plane transverse to
the trajectory. When the point on the track is the one with closest transverse
approach with respect to $(x_r,y_r)$: $\phi=\phi_0$, $x_\perp = \delta$ and $z_\perp =
z_0 \cos\lambda$. 

\end{itemize}

All equations presented in this report can be adapted to any of these alternative
parametrizations in a straightforward way. For completeness, we also collect in Appendix~A 
the Jacobians of the transformations to the alternative conventions. These Jacobians are
required for the propagation of the uncertainties discussed in Section~\ref{sec:changeref}.

\section{Some useful formulae for propagation algorithms in the zero-curvature limits \label{sec:answers}}

Most of the problems encountered in propagation algorithms admit
several solutions from a formal point of view. However, many of these
solutions are not optimal, neither from the point of view of precision,
nor from the point of view of simplicity.  The relations shown below
are free of divergences in the $C\rightarrow 0$ limit.

\subsection{Determining the azimuthal angle of the trajectory at a given point 
in the transverse plane. \label{sec:getphi}}

If the curvature is already available, an optimal expression to determine $\phi$ is the 
following:
\begin{eqnarray}
  \phi & = & \mathrm{atan2} \left( \sin\phi_0 - C~(x-x_0) ,
                      \cos\phi_0 + C~(y-y_0) \right ) \label{findphi1}
\end{eqnarray}

\noindent 
where $\mathrm{atan2}(y,x)$ is the function that returns the 
azimuthal angle of the vector $(x,y)$ in the 
range $(-\pi,\pi]$. If one prefers not to use $C$, an alternative expression 
is:
\begin{eqnarray}
 \phi & = & \mathrm{InRange}(2~\mathrm{atan2}(y-y_0,x-x_0)-\phi_0) \label{findphi2}
\end{eqnarray}

\noindent where $\mathrm{InRange}(\phi)$ is the function that 
brings the angle $\phi$ into the $(-\pi,\pi]$ range:
\begin{eqnarray}
 \mathrm{InRange}(\phi) & = & \left\{ 
\begin{array}{lrl}
      \phi+2\pi~\mathrm{int}(0.5-\frac{\phi}{2\pi}) & \hfil & {\rm, for}~\phi \le \pi \\
      \phi-2\pi~\mathrm{int}(0.5+\frac{\phi}{2\pi}) & \hfil & {\rm, for}~\phi > \pi
\end{array} 
\right.
\end{eqnarray}

Here $\mathrm{int}(x)$ is the integer part of the real number $x$.
Both Equations, \ref{findphi1} and~\ref{findphi2}, give a precise
answer in the $C \rightarrow 0$ limit.

\subsection{Curvature when only two points in the transverse plane and the azimuthal
angle at one of these points are known.}

  The optimal and shortest answer for this problem is:
\begin{eqnarray}
      C & = & \frac{2 (x-x_0)\sin\phi_0-2 (y-y_0)\cos\phi_0}
                   {(x-x_0)^2 + (y-y_0)^2}
\end{eqnarray}  

\noindent
where $\phi_0$ is the azimuthal angle at the point $(x_0,y_0)$ and $(x,y)$ is a second 
point on the trajectory.

\subsection{Arc length at a given point of the trajectory.}

If we define the following projected components of the distance between the point of closest transverse approach $(x_0,y_0)$ and the given point $(x,y)$:

\begin{eqnarray}
\Delta_\parallel & = & (x-x_0)\cos\phi_0 + (y-y_0)\sin\phi_0 \\
\Delta_\perp & = & - (x-x_0)\sin\phi_0 + (y-y_0)\cos\phi_0,  
\end{eqnarray}

\noindent
we obtain the solution, well behaved in the zero curvature limit:
\begin{eqnarray}
  s = \frac{\mathrm{atan2}(C~\Delta_\perp, 1 + C~\Delta_\parallel)}{C} \label{scalc}
\end{eqnarray}

If we have access to the local azimuthal angle $\phi$ at $(x,y)$ (see also 
subsection~\ref{sec:getphi}), we can use instead a relation that is 
totally independent of $C$:

\begin{eqnarray}
 s & = & \frac{\Delta_\parallel}
              {\mathrm{sinc} (\phi-\phi_0)} \label{scalc2}
\end{eqnarray}      

\noindent
where $\mathrm{sinc}(x)$ is the function $(\sin x)/x$.

\subsection{Determining the position on the trajectory for a given arc length.} 
              
To avoid accuracy problems one should use the curvature $C$ in a special way:
\begin{eqnarray}
  x & = & x_0 + s~\mathrm{sinc}\left(\frac{Cs}{2}\right) 
   \cos\left(\phi_0 - \frac{Cs}{2}\right) 
\label{xprime} \\ 
  y & = & y_0 + s~\mathrm{sinc}\left(\frac{Cs}{2}\right) 
   \sin\left(\phi_0 - \frac{Cs}{2}\right), 
\label{yprime}
\end{eqnarray}

If $\phi$ is available, one can make the substitution: $Cs = \phi_0 - \phi$, 
and use instead: 

\begin{eqnarray}
  x & = & x_0 - s~\mathrm{sinc}\left(\frac{\phi-\phi_0}{2}\right) 
   \cos\left(\frac{\phi+\phi_0}{2}\right) \\
  y & = & y_0 - s~\mathrm{sinc}\left(\frac{\phi-\phi_0}{2}\right) 
   \sin\left(\frac{\phi+\phi_0}{2}\right), 
\end{eqnarray}

\section{Propagation of track parameters and uncertainties \label{sec:changeref}}

\begin{figure}[ht]
\begin{center}
  \includegraphics[width=0.9\linewidth]{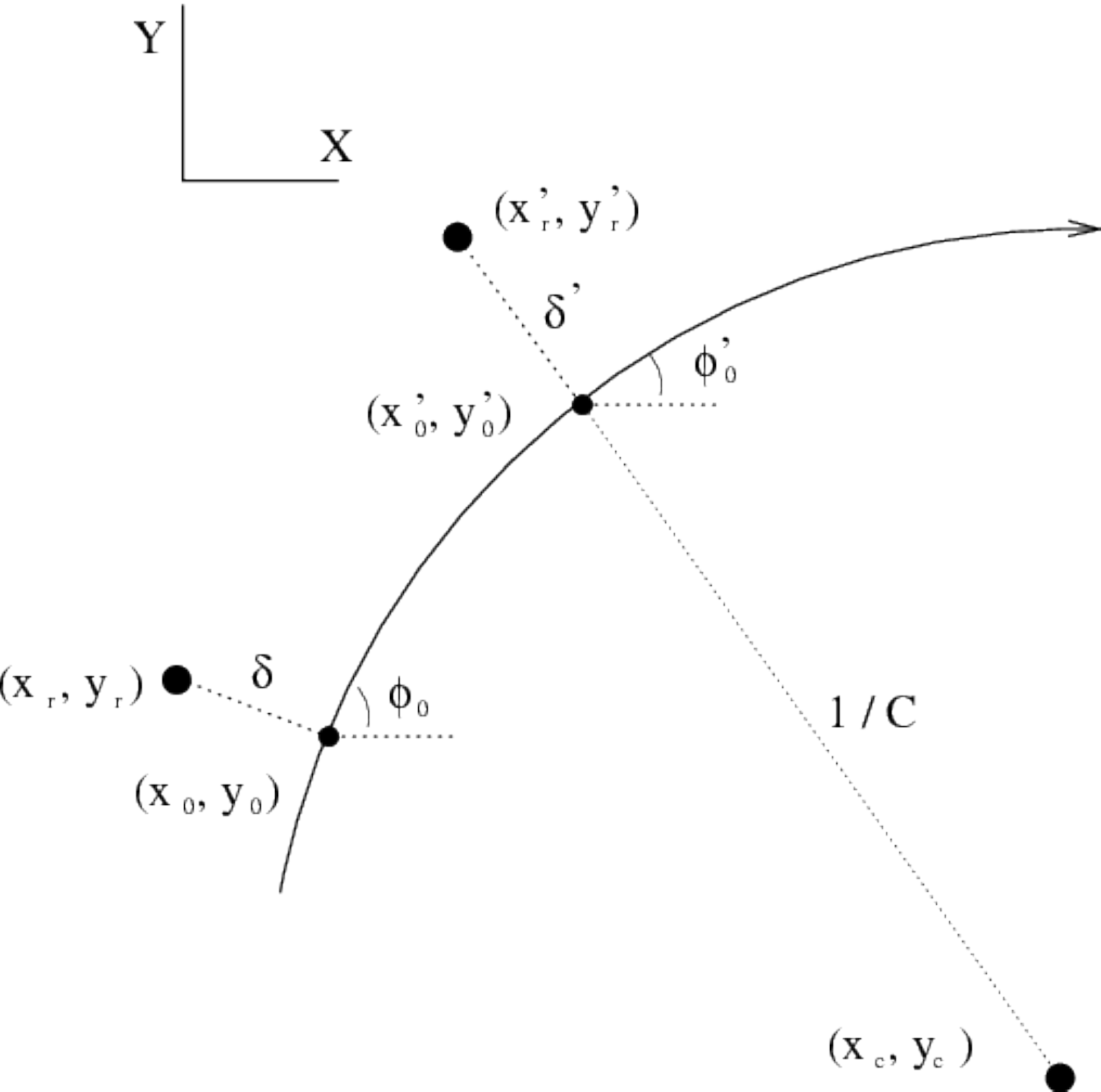}
\end{center}

\caption{ Change of parameters in XY when the reference point moves
from $(x_r,y_r)$ to $(x_r^\prime,y_r^\prime)$.}

\label{movereffig}
\end{figure}

In this section we provide expressions to update the track parameters
and the associated covariance matrix when a new reference point
$(x_r^\prime,y_r^\prime,z_r^\prime)$, different from the original one
$(x_r,y_r,z_r)$, is used in the parametrization
(Figure~\ref{movereffig}). There are two important use cases that are
covered with this exercise: a) a track referred to a new point that is
different from the initial estimate used in reconstruction (typically
the beam spot); b) the propagation to the crossing point between the
track and any given subdetector surface.

\subsection{Updated track parameters at a new reference point}

 Changing the reference point of a track implies a modification of
its parameters $\phi_0$, $\delta$ and $z_0$. We will denote the 
new parameters by $\phi_0^\prime$, $\delta^\prime$ and $z_0^\prime$.
Since $\mathrm{sign(1-C\delta) = sign(1-C\delta^\prime)}$ for all 
cases of practical interest, 
$\phi_0^\prime$ can be unambiguously determined as follows:
\begin{eqnarray}
  \phi_0^\prime =
 \mathrm{atan2} \left( \sin\phi_0 - \frac{C\Delta_x}{1-C\delta} ,
                      \cos\phi_0 + \frac{C\Delta_y}{1-C\delta} \right )
\label{phiprime}
\end{eqnarray}

\noindent
where $\Delta_x = x_r^\prime - x_r$ and $\Delta_y = y_r^\prime - y_r$.
Once $\phi_0^\prime$ is available, a convenient expression to determine 
$\delta^\prime$, also with a good behavior in the $C \rightarrow 0$ limit, is:
\begin{eqnarray}
  \delta^\prime & = & \delta 
 + \Delta_x\sin\phi_0-\Delta_y\cos\phi_0
 + (\Delta_x\cos\phi_0+\Delta_y\sin\phi_0)
    \tan(\frac{\phi_0^\prime-\phi_0}{2}) \label{dcaprime}
\end{eqnarray} 

   The point of minimum transverse approach to $(x_r^\prime,y_r^\prime)$ is 
then simply given by:
\begin{eqnarray}
   x_0^\prime & = & x_r^\prime - \delta^\prime \sin\phi_0^\prime \\
   y_0^\prime & = & y_r^\prime + \delta^\prime \cos\phi_0^\prime,
\end{eqnarray}

\noindent
leading to the following expression for the determination of the
$z_0^\prime$ parameter: 
\begin{eqnarray}
   z_0^\prime & = & z_0 + 
\frac{(x_0^\prime-x_0)\cos\phi_0 + (y_0^\prime-y_0)\sin\phi_0}
{\mathrm{sinc}(\phi_0^\prime-\phi_0)}~\tan\lambda
\end{eqnarray}

\subsection{Approximate solutions for the distance of closest approach to a given point}

Equation~\ref{dcaprime} can be well approximated in the $C\delta^\prime
\muchless 1$ limit, i.e. when the new reference point is very close to
the particle trajectory: 

\begin{eqnarray}
  \delta^\prime \approx \delta - \frac{C\delta^2}{2}
  + \left( \Delta_x \sin\phi_0 - \Delta_y \cos\phi_0 \right)
    ~( 1-C\delta )
  - C~\frac{\Delta_x^2+\Delta_y^2}{2} \label{crossing}
\end{eqnarray}

\noindent
where the neglected term on the right hand side is $C\delta^{\prime 2} / 2$. 
The expression can be further simplified if
$C\delta^2$ and $C\delta\delta^\prime$ terms are also negligible: 
\begin{eqnarray}
  \delta^\prime \approx \delta 
  + \Delta_x \sin\phi_0 - \Delta_y \cos\phi_0 
  - \frac{C}{2(1-C\delta)}~\left( \Delta_x^2+\Delta_y^2 \right)
\end{eqnarray}

\noindent
This approximation can be used as the starting point for the determination 
of the track parameters $C,\phi_0,\delta$~\cite{karimaki}. It is indeed one of the 
preferred methods when energy losses are not important, and it was 
successfully employed in several past high-energy physics experiments.

\subsection{Updated covariance matrix after propagation to a new point or surface}

 The new covariance matrix after propagation, $V^\prime$, can be determined in a 
linear approximation as:
\begin{eqnarray}
      V^\prime & = &
J\left(\frac{\partial(C^\prime,\phi_0^\prime,\delta^\prime,\tan\lambda^\prime,z_0^\prime)}{\partial(C,\phi_0,\delta,\tan\lambda,z_0)}\right)~V~J\left(\frac{\partial(C^\prime,\phi_0^\prime,\delta^\prime,\tan\lambda^\prime,z_0^\prime)}{\partial(C,\phi_0,\delta,\tan\lambda,z_0)}\right)^T
\end{eqnarray}

where $J$ is the Jacobian matrix of the
transportation, $J^{T}$ its transpose and $V$ is the original
covariance matrix. We will implicitly assume that the full covariance
matrix can be built separately in the XY and SZ projections. This
is a good approximation if: a) no significant correlations exist
between measurements in the XY and SZ planes or b) measurements in the
XY place are much more precise than measurements in the SZ plane.  Both
conditions are generally satisfied in high-energy particle detectors.
Condition b) is satisfied by design in most cases of interest because 
the best precision is always required in the plane that measures the 
transverse momentum,
i.e. XY.  Condition a) is satisfied when active detectors adopt an
axial geometry around accelerator beams. An exception is the case of
silicon microstrip detectors with slightly tilted
stereo-layers~\footnote{This tilt corresponds to a small rotation
around the $(\cos\phi_d,\sin\phi_d,0)$ direction, where $\phi_d$ is the
azimuthal angle of the center of the layer at $z=0$.}. In this case,
condition b) is approximately satisfied, the influence of SZ
parameter uncertainties on the XY fit can be neglected and the arc
lengths described in the XY plane can be considered as extremely
precise compared with Z uncertainties in the SZ straight line fit.

The Jacobian matrix can be therefore expressed as:
\begin{eqnarray}
J\left(\frac{\partial(C^\prime,\phi_0^\prime,\delta^\prime,\tan\lambda^\prime,z_0^\prime)}{\partial(C,\phi_0,\delta,\tan\lambda,z_0)}\right) & \equiv & 
\left(
\begin{array}{c|c|c|c|c}
\frac{\partial C'}{\partial C} & \frac{\partial C'}{\partial\phi_0} & \frac{\partial C'}{\partial\delta} & 0 & 0 \\
\frac{\partial\phi_0^\prime}{\partial C} & \frac{\partial\phi_0^\prime}{\partial\phi_0} & \frac{\partial\phi_0^\prime}{\partial\delta} & 0 & 0 \\
\frac{\partial\delta^\prime}{\partial C} & \frac{\partial\delta^\prime}{\partial\phi_0} & \frac{\partial\delta^\prime}{\partial\delta} & 0 & 0 \\
0 & 0 & 0 & \frac{\partial\tan\lambda^\prime}{\partial\tan\lambda} & \frac{\partial\tan\lambda^\prime}{\partial z_0} \\
0 & 0 & 0 & \frac{\partial z_0^\prime}{\partial\tan\lambda} & \frac{\partial z_0^\prime}{\partial z_0}
\end{array}
\right)
\end{eqnarray}

After careful processing of all terms with a potentially bad behavior
in the straight line limit we obtain for the XY sub-matrix associated
to XY measurements:

\begin{eqnarray}
\frac{\partial C'}{\partial C} & = & 1 \\
\frac{\partial C'}{\partial\phi_0} & = & 0 \\
\frac{\partial C'}{\partial\delta} & = & 0
\end{eqnarray}
\begin{eqnarray}
\frac{\partial\phi_0^\prime}{\partial C} & = &
  -\frac{1}{(1-C\delta^\prime)^2}~
(\Delta_x \cos\phi_0 + \Delta_y \sin\phi_0) \\
\frac{\partial\phi_0^\prime}{\partial\phi_0} & = &
   \left( \frac{1-C\delta}{1-C\delta^\prime} \right) + 
   \frac{C^2 (1-C\delta)}{2(1-C\delta^\prime)^2}~ 
(\delta^{\prime 2} - (\Delta_x+\delta\sin\phi_0)^2
                   - (\Delta_y-\delta\cos\phi_0)^2 ) \\
\frac{\partial\phi_0^\prime}{\partial\delta} & = &
  -\frac{C^2}{(1-C\delta^\prime)^2}~
(\Delta_x \cos\phi_0 + \Delta_y \sin\phi_0) \\
\frac{\partial\delta^\prime}{\partial C} & = &  
   \frac{1}{2 (1-C\delta^\prime)}~
(\delta^{\prime 2} - (\Delta_x+\delta\sin\phi_0)^2
                   - (\Delta_y-\delta\cos\phi_0)^2 ) \label{jacofirst} \\ 
\frac{\partial\delta^\prime}{\partial\phi_0} & = &
   \frac{1-C\delta}{1-C\delta^\prime}~
(\Delta_x \cos\phi_0 + \Delta_y \sin\phi_0) \\
\frac{\partial\delta^\prime}{\partial\delta} & = & 
1 + \frac{C^2}{2(1-C\delta^\prime)}~           
(\delta^{\prime 2} - (\Delta_x+\delta\sin\phi_0)^2
                   - (\Delta_y-\delta\cos\phi_0)^2 )  
\label{jacolast}
\end{eqnarray}

\noindent
with a smooth transition to the straight-line track limit:
\begin{eqnarray}
\frac{\partial C'}{\partial C} & = & 1 \\
\frac{\partial C'}{\partial\phi_0} & = & 0 \\
\frac{\partial C'}{\partial\delta} & = & 0 \\
\frac{\partial\phi_0^\prime}{\partial C} & = &
-(\Delta_x \cos\phi_0 + \Delta_y \sin\phi_0) \\
\frac{\partial\phi_0^\prime}{\partial\phi_0} & = & 1 \\
\frac{\partial\phi_0^\prime}{\partial\delta} & = & 0 \\
\frac{\partial\delta^\prime}{\partial C} & = & 
   \frac{1}{2}~
(\delta^{\prime 2} - (\Delta_x+\delta\sin\phi_0)^2
                   - (\Delta_y-\delta\cos\phi_0)^2 ) \\ 
\frac{\partial\delta^\prime}{\partial\phi_0} & = &
(\Delta_x \cos\phi_0 + \Delta_y \sin\phi_0) \\
\frac{\partial\delta^\prime}{\partial\delta} & = & 1 
\end{eqnarray}

The derivatives required to build the $(\tan\lambda,z_0)$ covariance 
sub-matrix are simpler:
\begin{eqnarray}
\frac{\partial\tan\lambda^\prime}{\partial\tan\lambda} & = & 1 \\
\frac{\partial\tan\lambda^\prime}{\partial z_0} & = & 0 \\
\frac{\partial z_0^\prime}{\partial\tan\lambda} & = & 
     s \\
\frac{\partial z_0^\prime}{\partial z_0} & = & 1
\end{eqnarray}

For other track parametrizations (see section~\ref{subsec:other_params}), 
an additional conversion 
will be necessary: $V_{par}^\prime = J_{par}~V_{par}~J_{par}^T$, where
$V_{par}$ is the covariance matrix of the alternative parametrization and 
$J_{par}$ is related with the Jacobian matrices given in Appendix A via the 
formula: 
\begin{eqnarray}
      J_{par} & = & J\left(\frac{\partial(\mathrm{new~par.})}{\partial(C,\phi_0,\delta,\ldots)}\right)~J\left(\frac{\partial(C^\prime,\phi_0^\prime,\delta^\prime,\ldots)}{\partial(C,\phi_0,\delta,\ldots)}\right)~J\left(\frac{\partial(\mathrm{new~par.})}{\partial(C,\phi_0,\delta,\ldots)}\right)^{-1}
\end{eqnarray}

\subsection{Track crossing a cylinder with axis parallel to Z \label{subsec:cylinder}}

\begin{figure}[ht]
\begin{center}
    \includegraphics[width=0.9\linewidth]{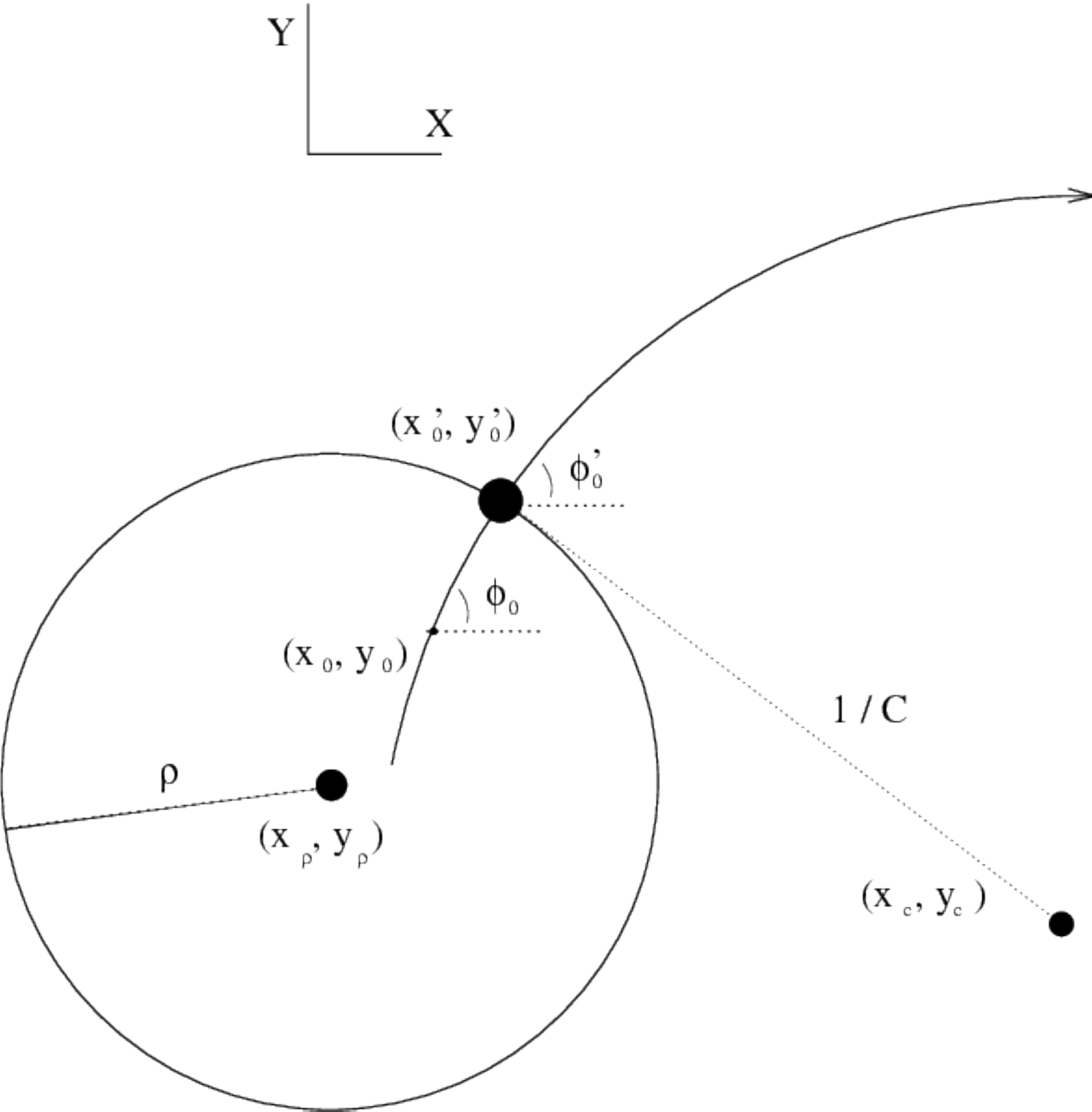}
\end{center}

\caption{XY projection of a track crossing a cylinder with axis parallel to the Z axis.}

\label{fig:crossc}
\end{figure}

   The case is illustrated in Figure~\ref{fig:crossc}. The crossing point 
$(x_0^\prime,y_0^\prime,z_0^\prime)$ is defined by the conditions:
\begin{eqnarray}
   C~x_0^\prime + \sin\phi_0^\prime & = & C~x_0 + \sin\phi_0 \\
   C~y_0^\prime - \cos\phi_0^\prime & = & C~y_0 - \cos\phi_0 \\
   z_0^\prime & = & z_0 + s~\tan\lambda \label{zprimeline} \\
   \rho^2 & = & (x_0^\prime-x_\rho)^2 + (y_0^\prime-y_\rho)^2
\end{eqnarray}

\noindent
where $(x_\rho,y_\rho)$ is the center of the circumference being crossed
in the XY plane
and $\rho$ is its radius. To solve the system of equations we use the following
definitions:
\begin{eqnarray}
    \phi_\rho & = & \mathrm{atan2} 
\left( \sin\phi_0-C (x_\rho-x_0), \cos\phi_0+C (y_\rho-y_0) \right) \\
    \phi_c & = & \mathrm{atan2} 
\left( y_0^\prime-y_\rho, x_0^\prime-x_\rho \right) \\
    \gamma & = & 
\frac{2(x_\rho-x_0)\sin\phi_0-2 (y_\rho-y_0)\cos\phi_0-C\rho^2
 -C((x_\rho-x_0)^2+(y_\rho-y_0)^2)}
{2 \rho \sqrt{(\sin\phi_0-C(x_\rho-x_0))^2+(\cos\phi_0+C(y_\rho-y_0))^2}}
\end{eqnarray}

   The implicit solution is simply: 
\begin{eqnarray}
    \sin(\phi_c-\phi_\rho) & = & \gamma
\end{eqnarray}

\noindent
which has a meaning only if $\mid \gamma \mid < 1$ and provides two possible solutions
for $\phi_c$ (as visually expected). In terms of $\phi_c$, the values of $x_0^\prime$, 
$y_0^\prime$, $\phi_0^\prime$ and $z_0^\prime$ are given by:
\begin{eqnarray}
    x_0^\prime & = & x_\rho + \rho~\cos\phi_c \\
    y_0^\prime & = & y_\rho + \rho~\sin\phi_c \\
    \phi_0^\prime & = & \mathrm{atan2} 
\left( \sin\phi_0-C (x_0^\prime-x_0), \cos\phi_0+C (y_0^\prime-y_0) \right) \\
    z_0^\prime & = & z_0 + \left[ \frac{(x_0^\prime-x_0)\cos\phi_0+(y_0^\prime-y_0)\sin\phi_0}{\mathrm{sinc} (\phi_0^\prime-\phi_0)}\right]~\tan\lambda
\end{eqnarray}

In most cases we will only be interested in solutions with a positive arc length:
\begin{eqnarray}
 (x_0^\prime-x_0) \cos\phi_0 + (y_0^\prime-y_0) \sin\phi_0 > 0
\end{eqnarray}

   In a typical scenario with a cylinder with $\rho^2 < R^2$ and 
$(x_0-x_\rho)^2+(y_0-y_\rho)^2 < \rho^2$, i.e. a cylinder with its 
center close to the distance of minimum approach, 
only one solution with positive arc length exists, given by:
\begin{eqnarray}
    \phi_c & = & \phi_\rho + \arcsin(\gamma)
\end{eqnarray}

The covariance matrix at the new crossing point 
can be determined from 
the equations developed in Section~\ref{sec:changeref},
using $(x_0^\prime, y_0^\prime,z_0^\prime)$
as the new reference point of the trajectory. Note that, by construction, the reference 
point is sitting on the trajectory and therefore $\delta^\prime=0$.

\subsection{Track crossing a plane \label{subsec:plane}}

This case is illustrated in Figure~\ref{fig:crossp}. The crossing point 
$(x_0^\prime,y_0^\prime,z_0^\prime)$ can be found from the following conditions:

\begin{figure}[ht]
\begin{center}
    \includegraphics[width=0.9\linewidth]{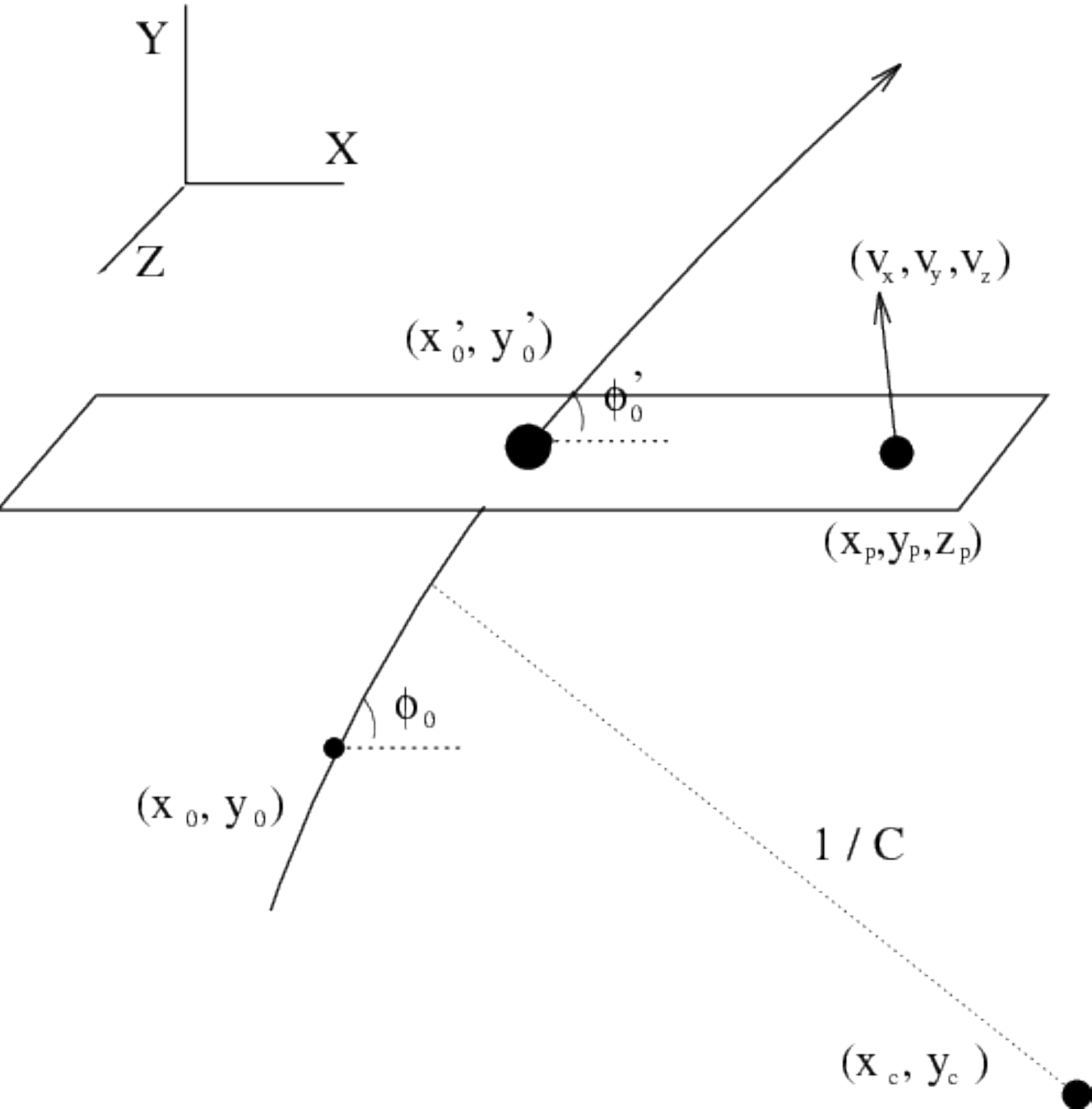}
\end{center}

\caption{Parameters relevant to find the point where a track crosses a plane. In the figure 
$(x_p,y_p,z_p)$ is a point in the plane and $(v_x,v_y,v_z)$ is a unitary vector 
perpendicular to it.}

\label{fig:crossp}
\end{figure}

\begin{eqnarray}
   x_0^\prime & = & x_0 + s~\mathrm{sinc}\left(\frac{Cs}{2}\right) 
    \cos\left(\phi_0 - \frac{Cs}{2}\right) \\
   y_0^\prime & = & y_0 + s~\mathrm{sinc}\left(\frac{Cs}{2}\right) 
    \sin\left(\phi_0 - \frac{Cs}{2}\right)
\end{eqnarray}
\begin{eqnarray}
   z_0^\prime & = & z_0 + s~\tan\lambda \\
   0 & = & (x_p-x_0^\prime)~v_x + (y_p-y_0^\prime)~v_y 
   + (z_p-z_0^\prime)~v_z
\end{eqnarray}

\noindent
where $(x_p,y_p,z_p)$ is a point in the plane and 
$(v_x,v_y,v_z)$ is a unitary vector perpendicular to it. To solve the 
problem we define:
\begin{eqnarray}
    \phi_v & = & \mathrm{atan2} \left( v_y,v_x \right) \\
    v_t & = & \sqrt{v_x^2+v_y^2} \\
    d_p & = & (x_p-x_0)~v_x + (y_p-y_0)~v_y + (z_p-z_0)~v_z
\end{eqnarray}

\noindent
where $d_p$ is actually the signed distance from $(x_0,y_0,z_0)$ to the plane.
The solution in terms of the variable $s$ can be obtained from 
following equation:
\begin{eqnarray}
  s & = & \frac{d_p}{v_t~\mathrm{sinc}\left(\frac{Cs}{2}\right)
  \cos\left(\frac{Cs}{2}+\phi_v-\phi_0\right) + 
  v_z~\tan\lambda} \label{eq:planeqn} 
\end{eqnarray}

\noindent
which makes sense only if $|Cs|<\pi$. From a practical point of view it is 
convenient to follow an iterative method converging to the solution:
\begin{eqnarray}
  s_0 & = & \frac{d_p}{v_t~\cos(\phi_v-\phi_0) + v_z~\tan\lambda} \\
  s_1 & = & \frac{d_p}{v_t~\mathrm{sinc}\left(\frac{Cs_0}{2}\right)
  \cos\left(\frac{Cs_0}{2}+\phi_v-\phi_0\right) + 
  v_z~\tan\lambda} \\
                 & \vdots & \nonumber \\
  s_i & = & \frac{d_p}{v_t~\mathrm{sinc}\left(\frac{Cs_{i-1}}{2}\right)
  \cos\left(\frac{Cs_{i-1}}{2}+\phi_v-\phi_0\right) + 
  v_z~\tan\lambda} 
\end{eqnarray}

Explicit solutions exist when $v_t=0$ or $v_z=0$. For $v_t=0$ 
(plane perpendicular to the Z axis): 
\begin{eqnarray}
  s & = & \frac{d_p}{v_z~\tan\lambda},
\end{eqnarray}

\noindent
whereas for the $v_z=0$ case (plane parallel to the Z axis):
\begin{eqnarray}
    \sin (Cs+\phi_v-\phi_0) & = & \sin(\phi_v-\phi_0) + C~d_p,
\end{eqnarray}

\noindent
which gives two possible solutions for $s$. Only the one with 
minimum arc length will be of interest in general. It is advisable
to determine the final $s$ value from $Cs$ not by division, but by 
substitution in Equation~\ref{eq:planeqn}, in order to keep the best 
possible accuracy.

\section{Including new points in a track~\label{sec:includepoint}}

   Let us consider a new measurement 
$(x_{meas},y_{meas},z_{meas})$ with some associated uncertainty. If the 
original track parameters and covariance matrix are known, it is always 
possible to improve the parameters of a track via $\chi^2$ methods.
According to the approach of Section~\ref{sec:changeref}, we will also 
assume that new measurement can be decomposed into uncorrelated 
measurements in the XY and SZ planes, such that XY and SZ fitting problems 
can be treated separately.

\subsection{XY plane}

Let us first redefine the track parameters using the new measurement
$(x_{meas},y_{meas},z_{meas})$ as a new reference point.
The new constraint can be interpreted as: $\delta=0 \pm \sigma$, 
where $\sigma$ is the transverse uncertainty on the measured position. 
The following $\chi^2$ can then be defined:
\begin{eqnarray}
\chi^2 & = & x_i~S_{ij}^0~x_j
       + \left( \frac{\delta}{\sigma} \right)^2 = x_i~S_{ij}^0~x_j
 	   + \left( \frac{\delta^{0} + x_3}{\sigma} \right)^2
\end{eqnarray}

\noindent
where $x_1 = C - C^0$, $x_2 = \phi_0 - \phi_0^0$, $x_3 = \delta - \delta^0$,
and a sum on repeated indices is assumed. $S_{ij}^0$ is the inverse of the 
original covariance matrix, $V_{ij}^0$, and $C^0$, $\phi_0^0$
and $\delta^0$ are the original values of the track parameters 
extrapolated to the new reference point. The 
minimum of this $\chi^2$ corresponds to:
\begin{eqnarray}
    V_{kj} & = & \left( S_{kj}^0 + \frac{1}{\sigma^2} g_{k3} g_{j3}
                       \right)^{-1} \\
& \Downarrow & \nonumber \\
 C & = & C^0 - V_{13} \frac{\delta^0}{\sigma^2} \\
 \phi_0 & = & \phi_0^0 - V_{23} \frac{\delta^0}{\sigma^2} \\
 \delta & = & \delta^0 - V_{33} \frac{\delta^0}{\sigma^2}
\end{eqnarray}

\noindent
where $g_{ij}$ is the identity matrix (1 if $i=j$, 0 otherwise).

\subsection{SZ plane}

The solution in this case is:
\begin{eqnarray}
 W_{ij} & = & \left( \begin{array}{cc} 
   T_{11}^0+ \frac{s^2}{\sigma^2}, & 
   T_{12}^0+ \frac{s}{\sigma^2} \\ 
   T_{21}^0+ \frac{s}{\sigma^2}, & 
   T_{22}^0+ \frac{1}{\sigma^2} \\ 
                     \end{array}  \right)^{-1} \\
& \Downarrow & \nonumber \\
 \tan\lambda & = & \tan\lambda^0 - (W_{11}~s+W_{12})~
           \frac{z_0^0 + s\tan\lambda-z_{meas}}{\sigma^2} \\
 z_0 & = & z_0^0 - (W_{21}~s+W_{22})~
           \frac{z_0^0+s\tan\lambda-z_{meas}}{\sigma^2}
\end{eqnarray}

\noindent
where $\tan\lambda^0$ and $z_0^0$ are the original SZ parameters of the 
track, $\sigma_z$ is the uncertainty on $z_{meas}$,
$s$ is the arc length described in the XY plane
and $T_{ij}^0$ is the inverse of $W_{ij}^0$, the original covariance matrix 
in the SZ plane. 

\subsection{Constraining a track to a point~\label{sec:fixpoint}}

   Sometimes we are not interested in a precise estimate of the uncertainties 
when a new measurement is added to a track, but only in the improvements 
obtained in the limit in which this measurement is much more precise than the 
uncertainty of the track extrapolation to it, i.e. the limit in which a 
track is constrained to a point. We will denote the new measurement 
by $(x_{fix},y_{fix},z_{fix})$.
In this case the equations presented in the previous section can be largely 
simplified.  

In the system in which the fixed point is the reference point, the 
solution for the XY plane when $\sigma\rightarrow 0$ is:
\begin{eqnarray}
   C & = & C^0 - \frac{V_{13}^0}{V_{33}^0}~\delta^0 \\ 
   \phi_0 & = & \phi_0^0
    - \frac{V_{23}^0}{V_{33}^0}~\delta^0 \\ 
   \delta & = & 0
\end{eqnarray}

\noindent
and a convenient parametrization of the covariance matrix in this limit is:
\begin{eqnarray}
 V_{ij} & = & \left( \begin{array}{ccc} 
\frac{S_{22}^0}{S_{11}^0 S_{22}^0-S_{12}^0 S_{12}^0} & 
\frac{-S_{12}^0}{S_{11}^0 S_{22}^0-S_{12}^0 S_{12}^0} & 0 \\
\frac{-S_{12}^0}{S_{11}^0 S_{22}^0-S_{12}^0 S_{12}^0} & 
\frac{S_{11}^0}{S_{11}^0 S_{22}^0-S_{12}^0 S_{12}^0} & 0 \\
                        0    & 0 & \sigma^2
                     \end{array}  \right)
\end{eqnarray}

\noindent
where the true (but small) value of the uncertainty $\sigma$ must be used to ensure that 
the covariance matrix is not singular. In the SZ plane we obtain: 
\begin{eqnarray}
   \tan\lambda & = & \tan\lambda^0
     - \frac{s W_{11}^0+W_{12}^0}{s^2 W_{11}^0 + 2 s W_{12}^0 + W_{22}^0}~
    (z_0^0+s\tan\lambda^0-z_{fix}) \\ 
   z_0 & = & z_{fix} -s \tan\lambda
\end{eqnarray}

\noindent
with a covariance matrix in the $\sigma_z \rightarrow 0$ limit given by: 
\begin{eqnarray}
 W_{ij} & = & \frac{1}{T_{11}^0-2 s T_{12}^0+s^2 T_{22}^0}
                    \left( \begin{array}{cc} 
                      1 + T_{22}^0\sigma_z^2, & -s-T_{12}^0\sigma_z \\ 
                      -s-T_{12}^0\sigma_z^2, & s^2+T_{11}^0\sigma_z^2 
                     \end{array}  \right)
\end{eqnarray}

\noindent
where again a non-null value of $\sigma_z$ is necessary to avoid 
potential singularities.

\section{Common vertex for several tracks in XY \label{sec:vertex_xy}}

Finding the common vertex for several tracks (at least 2) in the XY plane is equivalent to
finding a new common reference point, $(x_r^\prime,y_r^\prime)$ that globally minimizes the
distances of closest approach of all those tracks. We address the problem using
Equation~\ref{crossing} as starting point: 
 
\begin{eqnarray}
  \delta^\prime_i = \delta_i - \frac{C_i\delta^2_i}{2}
  + \left( \Delta_x \sin\phi_{0i} - \Delta_y \cos\phi_{0i} \right)
    ~( 1-C_i\delta_i )
  - C_i~\frac{\Delta_x^2+\Delta_y^2}{2} \label{eq:crossing0} 
\end{eqnarray}

\noindent
where $\delta^\prime_i$ is the distance of closest approach of the $ith$ track 
to the common vertex $(x_r^\prime,y_r^\prime)$. For convenience we 
choose to determine the parameters $\Delta_x$ and $\Delta_y$, which are the 
coordinates of the vector that connects the initial and final reference points: 
$(\Delta_x,\Delta_y) \equiv (x_r^\prime-x_r,y_r^\prime-y_r)$. 
Explicitly, the $\chi^2$ to be minimized is:
\begin{eqnarray}
   \chi_{xy}^2 & = & \sum_{i=1}^{N} \left( \frac{\delta^\prime_i}{\sigma_i} \right)^2
\end{eqnarray}

\noindent

where $\sigma_i$ is the error the distance of closest approach for the $ith$ track. 

The safest way to find the solution in terms of $\Delta_x$ and $\Delta_y$ the use standard
minimization programs like Minuit~\cite{minuit}. Nevertheless, if the initial reference
point $(x_r,y_r)$ is expected to be close to the final vertex, the solution can be found by
iteration using a Newton minimization method. The method is expected to converge as long as
$C_i~(\Delta_x^2+\Delta_y^2)$ terms stay small in the process. This is typically the case
of the determination of the primary vertex of the event, where the initial reference can be
well approximated by a nominal collision vertex position and the number of tracks is
sufficiently large to avoid the presence of far minima. Expanding the $\chi^2$ up to terms
quadratic in $\Delta_x, \Delta_y$: 

\begin{eqnarray}
   \alpha_i & = & \frac{2~\delta_i-C_i\delta^2_i}{2~\sigma_i} \\
   \beta_i & = & \frac{\sin\phi_{0i}~(1-C_i\delta_i)}{\sigma_i} \\
   \gamma_i & = & -\frac{\cos\phi_{0i}~(1-C_i\delta_i)}{\sigma_i} \\
   & \downarrow & \nonumber \\
   \chi_{xy}^2 & = & \sum_{i=1}^{N} \biggl[\biggr.
    \alpha^2_i + 
   \left(\begin{array}{cc} 2\alpha_i\beta_i, 2\alpha_i\gamma_i \end{array}\right)
    \left(\begin{array}{c} \Delta_x \\ \Delta_y \end{array}\right) \\
   & &  + 
   \left(\begin{array}{cc} \Delta_x, & \Delta_y \end{array}\right)
   ~\left(\begin{array}{cc}
   		\beta^2_i-C_i\alpha_i & \beta_i\gamma_i \\
   		\beta_i\gamma_i & \gamma^2_i-C_i\alpha_i \end{array}\right)~
    \left(\begin{array}{c} \Delta_x \\ \Delta_y \end{array}\right)
    + \ldots \biggl.\biggr]
\end{eqnarray}

The minimization of the quadratically truncated $\chi^2$ leads to the solution: 
\begin{eqnarray}
   \left( \begin{array}{c} \Delta_x \\ \Delta_y \end{array} \right) 
 & = & - ~\left(\begin{array}{cc}
   		      \beta^2_i-C_i\alpha_i & \beta_i\gamma_i \\
   		       \beta_i\gamma_i & \gamma^2_i-C_i\alpha_i 
   		 \end{array}\right)^{-1}
 ~ \left( \begin{array}{c} 
   \sum \alpha_i \beta_i \\ \sum \alpha_i \gamma_i \end{array} \right) 
\end{eqnarray}

This Newton step must be iterated until the vertex converges to a stable value. The
minimization can be stopped for instance when the distance between the vertices of two
consecutive iterations becomes of the order of the average impact parameter uncertainty:
$\approx (\sum 1/\sigma_i^2)^{-1/2}$. Note that, in each iteration, the reference point is
the vertex estimated in the previous iteration. Accordingly, the parameters $\phi_{0i}$ and
$\delta_i$ will have to be propagated to the new reference point using
Equations~\ref{phiprime} and~\ref{dcaprime} before the iteration starts. Let us point out
that, if convenient, the parameters of the tracks used in the vertex determination can be
further improved a posteriori by constraining them to the common vertex~\footnote{In
principle there is a remaining correlation effect between each track parameter and the 
common vertex position, which should be taken into account. However, this correlation is expected to be small as long as the total number of tracks is sufficiently large.}, using 
the equations developed in Section~\ref{sec:includepoint}. 

\section{Common vertex for several tracks in space \label{sec:vertex_sz}}

   The $\chi^2$ contribution in the SZ plane is:
\begin{eqnarray}
   \chi_z^2 & = & \sum_{i=1}^{N}
    \left( \frac{\Delta_Z - z_{0i} - 
  \frac{\tan\lambda_i}{\mathrm{sinc}(C_i s_i)}~
(\Delta_x\cos\phi_{0i}+\Delta_y\sin\phi_{0i})}{\sigma_{zi}} \right)^2
\end{eqnarray}

\noindent
where $s_i$ is the arc length from a displacement given by $(\Delta_x,\Delta_y)$ and
$\sigma_{zi}$ is the uncertainty on $z_{0i}$. Note that, due to the convention used in the
parametrization, $\Delta_Z$ is not a distance with respect to the reference point, but 
with respect to $z=0$. Again, $C \delta^\prime$ terms are neglected. The global $\chi^2$ to 
be minimized is: 

\begin{eqnarray}
  \chi^2 = \chi_{xy}^2 + \chi_{sz}^2 \;,
\end{eqnarray}

\noindent
which can be easily solved for $(\Delta_x,\Delta_y,\Delta_z)$ using standard minimization 
programs. Similarly to the $XY$ case, one can also try a Newton minimization procedure if 
the common vertex is not expected to be far from the initial reference point. 
In this 3D case, the solution of the quadratically truncated $\chi^2$ at each minimization 
step can be obtained from the following definitions: 

\begin{eqnarray}
   \alpha_i & = & \frac{2 \delta_i-C_i\delta^2_i}{2 \sigma_i} \\
   \beta_i & = & \frac{\sin\phi_{0i}~(1-C_i\delta_i)}{\sigma_i} \\
   \gamma_i & = & -\frac{\cos\phi_{0i}~(1-C_i\delta_i)}{\sigma_i} \\
   \xi_{i} & = & \frac{\cos\phi_{0i} \tan\lambda_i}
                     {\mathrm{sinc}(C_i s_{i})~\sigma_{zi}} \\
   \eta_{i} & = & \frac{\sin\phi_{0i} \tan\lambda_i}
                     {\mathrm{sinc}(C_i s_{i})~\sigma_{zi}} \\
   W_{\Delta x\Delta y\Delta z} & = & 
                    \left( \begin{array}{ccc} 
\sum (\beta_i^2+\xi_i^2-C_i\alpha_i) & \sum (\beta_i \gamma_i + \xi_i \eta_i) & 
-\sum \frac{\xi_i}{\sigma_{zi}} \\ 
& & \\
\sum (\beta_i \gamma_i + \xi_i \eta_i) & \sum (\gamma_i^2+\eta_i^2-C_i\alpha_i) &
-\sum \frac{\eta_i}{\sigma_{zi}} \\ 
& & \\
-\sum \frac{\xi_i}{\sigma_{zi}} & 
-\sum \frac{\eta_i}{\sigma_{zi}} & 
\sum \frac{1}{\sigma_{zi}^{2}} \end{array} \right)^{-1} \; ,
\end{eqnarray}

\noindent leading to the solution:

\begin{eqnarray}
   \left( \begin{array}{c} \Delta_x \\ \Delta_y \\ \Delta_z \end{array} \right) 
& = & - W_{\Delta x\Delta y\Delta z}~ \left( \begin{array}{c} 
   \sum (\alpha_i \beta_i + z_{0i} \frac{\xi_i}{\sigma_{zi}}) \\ \\
   \sum (\alpha_i \gamma_i + z_{0i} \frac{\eta_i}{\sigma_{zi}}) \\ \\
   -\sum \frac{z_{0i}}{\sigma_{zi}^2} \end{array} \right) 
\end{eqnarray}

Beside $\phi_{0i}$ and $\delta_i$, now also $s_i$ will have to be propagated to the 
estimated vertex from $\chi^2$ truncation before starting a new iteration. This can be done 
using Equations~\ref{scalc} or~\ref{scalc2}.

\bibliographystyle{myutphys}
\bibliography{helix_ciemat}

\providecommand{\href}[2]{#2}\begingroup\raggedright\begin{thebibliography}{10}

\bibitem{l3note}
J.~Alcaraz~Maestre, \textit{{Helicoidal tracks}}, L1 Internal Note 1666, 1994,
  \url{https://cern.ch/l3/note/notes1994.html}.

\bibitem{L3}
{L3}, B.~Adeva {\em et~al.}, \textit{{The Construction of the L3 Experiment}},
  \href{http://dx.doi.org/10.1016/0168-9002(90)90250-A}{{\em Nucl. Instrum.
  Meth. A} {\bfseries 289} (1990) 35--102}.

\bibitem{smd}
{L3 SMD}, M.~Acciarri {\em et~al.}, \textit{{The L3 silicon microvertex
  detector}}, \href{http://dx.doi.org/10.1016/0168-9002(94)91357-9}{{\em Nucl.
  Instrum. Meth. A} {\bfseries 351} (1994) 300--312}.

\bibitem{tec}
F.~Beissel {\em et~al.}, \textit{{Construction and performance of the L3
  central tracking detector}},
  \href{http://dx.doi.org/10.1016/0168-9002(93)90740-9}{{\em Nucl. Instrum.
  Meth. A} {\bfseries 332} (1993) 33--55}.

\bibitem{fruhwirth}
R.~Fruhwirth, \textit{{Application of Kalman filtering to track and vertex
  fitting}}, \href{http://dx.doi.org/10.1016/0168-9002(87)90887-4}{{\em Nucl.
  Instrum. Meth. A} {\bfseries 262} (1987) 444--450}.

\bibitem{Saxon}
J.~C. Hart and D.~H. Saxon, \textit{{Track and Vertex Fitting in an
  Inhomogeneous Magnetic Field}},
  \href{http://dx.doi.org/10.1016/0167-5087(84)90291-6}{{\em Nucl. Instrum.
  Meth. A} {\bfseries 220} (1984) 309--326}.

\bibitem{amspaper}
J.~Alcaraz, \textit{{Track fitting in slightly inhomogeneous magnetic fields}},
  \href{http://dx.doi.org/10.1016/j.nima.2005.07.021}{{\em Nucl. Instrum. Meth.
  A} {\bfseries 553} (2005) 613--619},
  \href{http://arxiv.org/abs/physics/0507129}{{\ttfamily
  arXiv:physics/0507129}}.

\bibitem{CMS}
{CMS}, S.~Chatrchyan {\em et~al.}, \textit{{The CMS Experiment at the CERN
  LHC}}, \href{http://dx.doi.org/10.1088/1748-0221/3/08/S08004}{{\em JINST}
  {\bfseries 3} (2008) S08004}.

\bibitem{Sijin}
P.~Billoir and S.~Qian, \textit{{Fast vertex fitting with a local
  parametrization of tracks}},
  \href{http://dx.doi.org/10.1016/0168-9002(92)90859-3}{{\em Nucl. Instrum.
  Meth. A} {\bfseries 311} (1992) 139--150}.

\bibitem{Salzburger}
A.~Salzburger, {\em {Track Simulation and Reconstruction in the ATLAS
  experiment}}, PhD thesis, Innsbruck U., 2008,
  \url{http://cds.cern.ch/record/2224514}.

\bibitem{ATLAS}
{ATLAS}, G.~Aad {\em et~al.}, \textit{{The ATLAS Experiment at the CERN Large
  Hadron Collider}},
  \href{http://dx.doi.org/10.1088/1748-0221/3/08/S08003}{{\em JINST} {\bfseries
  3} (2008) S08003}.

\bibitem{GEANE}
V.~Innocente, M.~Maire, and E.~Nagy, \textit{{GEANE: Average tracking and error
  propagation package}}, in {\em {Workshop on Detector and Event Simulation in
  High-energy Physics (MC '91)}}, pp.~58--78, 1991,
  \url{https://cds.cern.ch/record/247710}.

\bibitem{karimaki}
V.~Karimaki, \textit{{Effective circle fitting for particle trajectories}},
  \href{http://dx.doi.org/10.1016/0168-9002(91)90533-V}{{\em Nucl. Instrum.
  Meth. A} {\bfseries 305} (1991) 187--191}.

\bibitem{minuit}
F.~a. James, \textit{{MINUIT: Function Minimization and Error Analysis
  Reference Manual}}, \url{https://cds.cern.ch/record/2296388}, CERN Program
  Library Long Writeups.

\end{thebibliography}\endgroup

\section*{Appendix A. Alternative track parametrizations.}

  Here we collect the Jacobians 
$J\left(\frac{\partial(\mathrm{new~parameters})}{\partial(C,\phi_0,\delta,\tan\lambda,z_0)}\right)$ 
for the alternative parametrizations discussed in 
subsection~\ref{subsec:other_params}:

\begin{itemize}

\item Perigee parametrization~\cite{Sijin}:

\begin{eqnarray}
      \rho & \equiv & - C \\
      \phi_p & \equiv & \phi_0 \\
      \epsilon & \equiv & -\delta \\
      \theta & \equiv & \frac{\pi}{2} - \lambda \\
      z_p & \equiv & z_0 
\end{eqnarray}

\begin{eqnarray}
  J\left(\frac
              {\partial(\rho,\phi_p,\epsilon,\theta,z_p)}
              {\partial(C,\phi_0,\delta,\tan\lambda,z_0)}
   \right) = 
\left(
\begin{array}{c|c|c|c|c}
-1 & 0 & 0 & 0 & 0 \\ \hline
 0 & 1 & 0 & 0 & 0 \\ \hline
 0 & 0 & -1 & 0 & 0 \\ \hline
 0 & 0 & 0 & -\cos^2\lambda & 0 \\ \hline
 0 & 0 & 0 & 0 & 1
\end{array}
\right)
\end{eqnarray}

\item Modified perigee parametrization~\cite{Salzburger}:

\begin{eqnarray}
      \frac{q}{p} & \equiv & \frac{C \cos\lambda}{0.29979~B} \\
      \phi_p & \equiv & \phi_0 \\
      \epsilon & \equiv & -\delta \\
      \theta & \equiv & \frac{\pi}{2} - \lambda \\
      z_p & \equiv & z_0 
\end{eqnarray}

\begin{eqnarray}
  J\left(\frac
              {\partial(q/p,\phi_p,\epsilon,\theta,z_p)}
              {\partial(C,\phi_0,\delta,\tan\lambda,z_0)}
   \right) =
\left(
\begin{array}{c|c|c|c|c}
\frac{\cos\lambda}{0.29979~B} & 0 & 0 & -\frac{C\sin\lambda\cos^2\lambda}{0.29979~B} & 0 \\ \hline
 0 & 1 & 0 & 0 & 0 \\ \hline
 0 & 0 & -1 & 0 & 0 \\ \hline
 0 & 0 & 0 & -\cos^2\lambda & 0 \\ \hline
 0 & 0 & 0 & 0 & 1
\end{array}
\right) 
\end{eqnarray}

\item Curvilinear parametrization at the minimum distance of transverse approach~\cite{GEANE}:

\begin{eqnarray}
      \frac{q}{p} & \equiv & \frac{C \cos\lambda}{0.29979~B} \\
      \phi & \equiv & \phi_0 \\
      x_\perp & \equiv & \delta \\
      \lambda & & \textrm{unchanged} \\
      z_\perp & \equiv & z_0~\cos\lambda 
\end{eqnarray}

\begin{eqnarray}
  J\left(\frac
              {\partial(q/p,\phi,x_\perp,\lambda,z_\perp)}
              {\partial(C,\phi_0,\delta,\tan\lambda,z_0)}
   \right) =  
\left(
\begin{array}{c|c|c|c|c}
\frac{\cos\lambda}{0.29979~B} & 0 & 0 & -\frac{C\sin\lambda\cos^2\lambda}{0.29979~B} & 0 \\ \hline
 0 & 1 & 0 & 0 & 0 \\ \hline
 0 & 0 & 1 & 0 & 0 \\ \hline
 0 & 0 & 0 & -\cos^2\lambda & 0 \\ \hline
 0 & 0 & 0 & -z_0\sin\lambda\cos^2\lambda & \cos\lambda
\end{array}
\right)
\end{eqnarray}

\end{itemize}

\end{document}